\documentclass[twocolumn]{aastex63}

\usepackage{graphicx}
\usepackage{amsmath}	
\usepackage{amssymb}
\usepackage{dcolumn}
\usepackage{color}
\usepackage{bm}
\usepackage{newtxtext,newtxmath}
\usepackage[T1]{fontenc}
\usepackage{inputenc}
\usepackage{ae,aecompl}
\usepackage{natbib}
\usepackage{blindtext}

\usepackage{hyperref} 
\usepackage{breakurl}

\submitjournal{ApJ}

\shorttitle{Cosmic ray  halos}
\shortauthors{Recchia et al.}

\begin{document}

\title{Giant cosmic ray halos around M31 and the Milky Way}

\correspondingauthor{Sarah Recchia}
\email{sarah.recchia@unito.it}

\author{S. Recchia}
\affiliation{Dipartimento di Fisica, Universit\'a di Torino, via P. Giuria 1, 10125 Torino, Italy}
\affiliation{Istituto Nazionale di Fisica Nucleare, Sezione di Torino, Via P. Giuria 1, 10125 Torino, Italy}
\affiliation{Universit\'e de Paris, CNRS, Astroparticule et Cosmologie, F-75006 Paris, France
}
\affiliation{IJCLab, CNRS/IN2P3, Université Paris-Saclay, F-91405 Orsay, France}

\author{S. Gabici}
\affiliation{Universit\'e de Paris, CNRS, Astroparticule et Cosmologie, F-75006 Paris, France
}
\author{F. A. Aharonian}
\affiliation{Dublin Institute for Advanced Studies, 31 Fitzwilliam Place, Dublin 2, Ireland}
\affiliation{Max-Planck-Institut f\"ur Kernphysik, Postfach 103980, D-69029 Heidelberg, Germany}

\author{V. Niro}
\affiliation{Universit\'e de Paris, CNRS, Astroparticule et Cosmologie, F-75006 Paris, France
}

\nocollaboration{4}



\begin{abstract}

Recently, a diffuse emission of 1-100 GeV $\gamma$-rays has been detected from the direction of Andromeda.
The emission is centered on the galaxy, and extends for $\sim 100-200$ kpc away from its center.
Explaining the extended $\gamma-$ray emission within the framework of standard scenarios for the escape of cosmic rays  injected in the galactic disk or in the galactic center is problematic.
In this paper, we argue that a cosmic ray origin (either leptonic or hadronic) of the $\gamma$-ray emission is  possible in the framework of non standard cosmic ray propagation scenarios or in the case of particle acceleration taking place in the galaxy's halo. It would imply the existence of a giant cosmic ray halo surrounding M31, possibly powered by the galaxy nuclear activity, or by accretion of intergalactic gas.
Remarkably, if cosmic ray halos, as the one observed around M31, are a common feature of galaxies, including our own, the interactions between cosmic ray protons and the Milky Way circumgalactic gas could also explain the isotropic diffuse flux of neutrinos observed by Icecube.
\end{abstract}




\section{Introduction}
\label{sec:intro}

The Andromeda Galaxy (M31), located at a distance of $\sim 785$ kpc,  is the closest spiral galaxy to the Milky Way (MW), and with the MW shares many similarities. 
Both galaxies are composed of a bulge, a disk, an extended gaseous halo, a central supermassive black hole and a dark matter halo which extends for $~200-300$ kpc (in radius) with a total mass of $\sim 10^{12}M_{\odot}$ (see ~\cite{Moskalenko-gamma-M31}, and references therein). 

Attempts to detect M31 in $\gamma-$rays date back to the seventies~\citep{Fichtel-1975, Pollock-1981, Sreekumar-1994, Hartman-1999}, but high energy photons were observed only thank to the \textit{Fermi}-LAT experiment~\citep{Abdo-2010, Ogelman-2011,Ackermann-2017, DiMauro-2019-DM-M31}. 
The integrated gamma-ray luminosity above 100 MeV was found to be $\sim 6.6 \times 10^{41}$ s$^{-1}$, very close (within less than a factor of 2) to that of the MW~\citep{Abdo-2010}.
While for the MW the emission correlates spatially with the gaseous disk, for M31 it appears to be concentrated
within the inner $\sim$ 5 kpc region~\citep{Ackermann-2017}.
The origin of the emission remains debated.
Finally, some evidence for the existence of structures similar to the Fermi Bubbles emanated from the central region of M31 has also been reported~\citep{Pshirkov-2016}.

Recently, an analysis of \textit{Fermi}-LAT data revealed the presence of an extended $\gamma-$ray emission from a  very large halo surrounding Adromeda~\citep{Moskalenko-gamma-M31}.
The authors investigated a region of $28^{\circ}\times 28^{\circ}$ which includes a projected radius of $\sim 200$ kpc from the center of M31.
After performing an accurate modeling of the MW foreground emission, they found  an excess that extends up to about $\sim 120-200$ kpc around the center of M31. In order to better characterize such emission, they included in the analysis a spherically symmetric template, centered on M31 and further divided into three regions: the inner galaxy (IG), a region of $\sim 5.5$ kpc radius  that contains the bright $\gamma$-ray emission from the inner galaxy~\citep{Ackermann-2017}, the spherical halo (SH), an intermediate ring which extends up to $\sim 120 $ kpc and the outer halo (OH), a ring of  $\sim 120-200$ kpc. 
The authors concluded that the excess emission  comes indeed from  M31,  and estimated the total $\gamma-$ray flux and spectrum in the three regions. 

Since the northern part of the considered  regions, and especially that of the OH ring, partially overlaps with the disk of the MW, the authors also performed an additional analysis in which they reported separately the total flux and spectrum of the north and south part of the intermediate (SH)  and outer (OH) rings. While the north/south regions of the SH do not show relevant spectral differences, in the case of the OH the two spectra are quite different, with a bumpy profile in the northern part, showing  a very likely contamination from the MW.
Therefore, the authors conclude that, while the excess from the SH region is likely associated to the halo of M31, that from the OH region has a less clear origin, and could be partly or completely related to the MW, or even have another unspecified origin.

Based on  the large extension of the emitting region, on the spectral shape and on the intensity of the various components, the authors of~\cite{Moskalenko-gamma-M31} suggest that, while some fraction of the $\gamma-$ray emission could be due to cosmic ray (CR) interactions in the halo of M31, it is unlikely that such CRs may dominate the production of  $\gamma-$rays. Instead they suggested,  that a dark matter interpretation could be a better explanation, and described the details of such interpretation in a recent publication~\citep{moskalenkodarkmatter}. 

As we argue in this paper, a CR origin for the extended emission is not only possible, but even quite natural. It would imply the existence of a giant CR halo of radius $\sim 100$ kpc surrounding Andromeda, and would require a non standard scenario for the transport of CRs into galactic halos.
Remarkably, the existence of such large halos was proposed for both the MW and M31~\citep{feldmann2013,Taylor-2014,Profumo-2020}. In particular, it was shown that the interaction of CRs in the diluted circumgalactic gas around the MW could explain the diffuse flux of neutrinos revealed by Icecube~\citep{Taylor-2014}, and a subdominant fraction of the isotropic gamma-ray background~\citep{feldmann2013}.
The presence of a CR halo surrounding both the MW and M31 would support the similarities in the non-thermal properties of the two galaxies.

The outline of the paper is the following: in Sec.~\ref{sec:data} we summarize the relevant results of the analysis of \textit{Fermi}-LAT from M31 reported by~\cite{Moskalenko-gamma-M31}; in Sec.~\ref{sec:energy}  we estimate the energy requirements for the hadronic or the leptonic origin of the $\gamma$-ray emission from the SH; 
in Sec.~\ref{sec:buoyant} we illustrate a scenario where CRs are produced during episodes of activity in the galactic center of M31 and then transported into the halo;
in Sec.~\ref{sec:acceleration} we  analyze a scenario where CRs are accelerated {\it in situ}, at a gigantic shock located in the SH; in  Sec.~\ref{sec:PeV} we explore the multi-wavelength and multi-messenger implications of a possible similarity between the halos surrounding M31 and the MW. A natural implication of this scenario is that Icecube neutrinos are originated in the extended halo of the MW. 
In Sec.~\ref{sec:multimessenger} we discuss the  implications of the existence of giant CR halos around more distant galaxies, and compute the expected signals in multi-TeV/PeV neutrinos and $\gamma-$rays from CR proton-proton interactions, along with the associated synchrotron emission from secondary electrons. We also discuss briefly the case of the galaxy NGC 1068~\citep{aartsen2020}; in Sec.~\ref{sec:concl} we draw our conclusions.\\

\section{Summary of the relevant $\gamma$-ray data}
\label{sec:data}

In this Section we summarize the results of the analysis of \textit{Fermi}-LAT from the M31 region performed in~\cite{Moskalenko-gamma-M31}.
We will not discuss here the bright $\gamma$-ray emission from the IG, whose origin has been debated in~\citep{Ackermann-2017,DiMauro-2019-DM-M31,profumo}, nor the tenuous diffuse emission from the OH, whose origin might be unrelated to M31~\citep{Moskalenko-gamma-M31}, but we will rather focus on the $\gamma$-ray emission observed from the SH.

At the distance of M31, the radial extension of the SH  with respect to the center of M31 is $5.5 ~{\rm kpc} \lesssim r \lesssim  120$ kpc. It corresponds to a solid angle of $3.42 \times 10^{-2}$ sr.
A spectral fit to the $\gamma-$ray emission observed from the SH was provided in~\cite{Moskalenko-gamma-M31}, where a power-law plus exponential cut-off parametrization was adopted:
\begin{equation}\label{eq:IE-generic}
I_{SH} \approx 9.8 \times 10^{-11}\,E_{\rm GeV}^{-1.9}\,e^{-E_{\rm GeV}/11.6} \rm MeV^{-1}\, cm^{-2}s^{-1}\, sr^{-1}
\end{equation}
where $E_{\rm GeV}$ is the photon energy in GeV (see Fig.~\ref{fig:fit-gamma}).
In order to minimize the contamination from the MW, we consider only $\gamma$-ray data from  the southern part of the SH.
Also a pure power law fit to the $\gamma$-ray emission from the (entire) SH was provided in~\cite{Moskalenko-gamma-M31}.
The results of the fit are shown in Fig.~\ref{fig:fit-gamma} and were obtained using the Fermi Science Support Center Interstellar Emission Models (FSSC IEM).

The total $\gamma$-ray luminosity of the entire (northern plus southern part) SH equal to:
\begin{equation}\label{eq:L-SH}
     L_{\gamma} \approx 1.7-1.9\times 10^{39}  \rm erg/s ~ .
\end{equation}
where the lower value corresponds to the power law plus cutoff fit to data, and the upper value to the pure power law case.
In the following, we will consider both fits, and we will discuss the profound implications of the presence or absence of a cutoff in the spectrum.

\section{Hadronic and leptonic origin of the $\gamma$-ray emission: energetics}
\label{sec:energy}

In this section we discuss a scenario where the gamma-ray emission from the halo of M31 is produced by CR interactions, either hadronic or leptonic. 
In the former case, $\gamma-$rays are due to  the decay on neutral pions produced in proton-proton interaction of CR protons with the diluted background gas in the halo, while in the latter they are produced via  inverse Compton scattering of CR electrons on CMB photons. 
Notice that synchrotron losses on the background magnetic field are most likely negligible at distances exceeding few tens of kiloparsecs from the disk, since the magnetic filed strength there is expected to be well below $\rm\sim 3~ \mu G$, and therefore its energy density to be subdominant with respect to that of the CMB~\citep{stanev-1997, Ferriere-2001, Jansson-2012}. 
In the following, under the assumption of stationarity, we compute the CR luminosity needed to explain the $\gamma$-ray emission from the SH.

\subsection{Leptonic scenario}
\label{sec:leptonic}

In this scenario, the $\gamma$-ray emission from the halo is due to inverse Compton scattering (ICS) of relativistic electrons off CMB photons. The average energy of such photons is $\langle \epsilon_{CMB} \rangle \sim 6.3 \times 10^{-4}$ eV. After the scattering, the photons are boosted to an energy~\citep{Blumenthal-1970}:
\begin{equation}
E_{\gamma} = \frac{4}{3} \gamma^2 \langle \epsilon_{CMB} \rangle \sim 3.2 ~E_{\rm TeV}^2 ~\rm GeV
\end{equation}
where $\gamma$ is the electron Lorentz factor and $E_{\rm TeV} = (E_e/1$ TeV) its energy in TeV.

This implies that the diffuse $\gamma$-ray emission seen by \textit{Fermi}-LAT at photon energies in the range $\sim 1-100$ GeV would be produced by electrons of energy $\sim 0.6-6$ TeV. 
The cutoff energy in the $\gamma$-ray spectrum ($E_{\gamma} = 11.6$ GeV) would correspond to electrons of energy  $E_{max,e} \approx 1.9$ TeV.

The inverse Compton energy loss rate for electrons in the CMB is given by~\citep{Blumenthal-1970}:
\begin{equation}
\label{eq:ICSlossrate}
\frac{{\rm d}E_e}{{\rm d}t} = \frac{4}{3} \sigma_T c \gamma^2 \omega_{CMB} \sim \, 2.5 \times 10^{-2} E_{\rm TeV}^2\, \rm  eV/s,
\end{equation}
where $\omega_{CMB} = 0.25\, \rm eV cm^{-3}$ is the energy density of CMB photons. 
The corresponding energy loss time is
\begin{equation}\label{eq:lossT-sync}
\tau_{CMB} \equiv \frac{E_e}{{\rm d}E_e/{\rm d}t} \sim 1.3\times 10^6  E_{\rm TeV}^{-1} ~\rm yr.
\end{equation}
This expression is valid in the Thomson limit, while at energies above $\sim 10$ TeV the Klein-Nishina effects becomes important~\citep{Blumenthal-1970}. 
This quite short timescale makes it very unlikely that such electrons may originate in the galactic disk or from the galactic center.
Even in the case of rectilinear (ballistic) motion, 3 TeV electrons  would move at most $\approx 100$ kpc before cooling. But for any realistic diffusion such distance would be much smaller.
Moreover, near the disk the energy loss time would be even shorter, due to the larger value of both the ambient magnetic field and the background photon field (starlight radiation plays a relevant role close to the disk). 
For these reasons, if the emission in the SH of M31 is of leptonic origin, the parent CR electron population is most likely  accelerated {\it in situ}.

The minimal energy requirement for this leptonic scenario can be estimated by assuming that the age of the system and/or the residence time of electrons in the acceleration region  is larger than the inverse Compton energy loss time.
Under these circumstances, the $\gamma-$ray production happens in a calorimetric regime, namely the observed $\gamma$-ray  luminosity (Eq.~\ref{eq:L-SH}) equals the electron luminosity:
\begin{equation}
\label{eq:Le}
L_e = L_{\gamma} \approx 1.7-1.9\times 10^{39} \rm erg/s
\end{equation}
Notice that this estimate refers only to electrons in the energy band $\approx 0.6-6$ TeV. 
Such a luminosity is of the same order than the total estimated power of CR electrons accelerated in the disk of the MW~\citep{strongMW}.

A qualitative fit to the $\gamma-$ray flux is shown in Fig.~\ref{fig:fit-gamma}, which has been obtained assuming a CR electron spectrum of  the form $\propto E_{\rm GeV}^{-2.0}\, e^{-E_{\rm GeV}/500}$. The ICS flux has been computed following~\cite{Khangulyan-2014-IC}. 

\subsection{Hadronic scenario}

In the hadronic scenario $\gamma-$rays of energy $E_{\gamma}$ are produced in proton-proton interactions between CR protons of energy $E_p \approx 10\,E_{\gamma}$ and the ambient gas~\citep{Kelner-2006-pp}. 
Moreover, the CR proton luminosity $L_p$ needed to explain the observed $\gamma$-ray luminosity $L_{\gamma}$ is given by~\citep{Taylor-2014}:
\begin{align}\label{eq:L-pp-general}
   L_p = \frac{3\, L_{\gamma}}{\mathit{f}}\\ \nonumber
   \mathit{f}= 1-e^{-\tau_{res}/\tau_{pp}},
\end{align}
where $\tau_{res}$ is the residence time of CRs in the halo and $\tau_{pp}$ is the timescale for pp interactions, 
\begin{equation}\label{eq:Tloss-pp}
\tau_{pp} \sim 7.1\times 10^{10} n_{H,-3}^{-1} \rm~ yr
\end{equation}
where $n_{H,-3}$ is the halo hydrogen density in units of $10^{-3}$ cm$^{-3}$.
The energy loss time $\tau_{pp}$ has been computed for a cross section $\sigma_{pp} \sim 3 \times 10^{-26}$ cm$^2$ and an inelasticity of the process $\kappa \approx 0.5$.
Note that for $n_{H,-3} \lesssim 1$ the energy loss time largely exceeds the age of the Universe.
Typical values of $ n_H$ expected at $\sim 100$ kpc from the disk of the MW  are $ \sim 10^{-4}-10^{-3} \rm cm^{-3}$~\citep{Miller-2013, Miller-2015}, and therefore CR protons don't lose energy. 
Such gas densities are consistent with a number of observational measurements~\citep{gupta2012}, which were recently confirmed in~\cite{Bridge-MW-M31-2020}, where the authors found some evidence for the existence of a local hot bridge, i.e. a cylinder of radius $\approx$ 120 kpc filled with hot gas connecting M31 to the MW.

Taking $\tau_{res} = 10^{9} \tau_{res,9}$ yr 
one can compute the CR proton luminosity needed to explain the $\gamma$-ray observations of the SH in M31:
\begin{equation}\label{eq:Lpp-SH}
L_{p} \approx 1.8\times 10^{41}\,  \tau_{res, 9}^{-1}\,n_{H, -3}^{-1} \rm erg/s\,.
\end{equation}
where a correction factor of $\approx 2$ has been applied in order to account for the enhancement of the $\gamma$-ray emission due to the presence of heavy nuclei in both CR and ambient gas~\citep{mori1997,caprioli2011,kafexhiu2014}.
Notice that for $\tau_{res, 9} \gtrsim 1$ and $n_{H,-3} \sim 1$, the luminosity of CR protons needed to account for the $\gamma$-ray emission from the SH 
is of the same order of that invoked to explain the population of CR protons observed in the disk of the MW~\citep{strongMW}.

\begin{figure}
	\centering
	\includegraphics[width=\columnwidth]{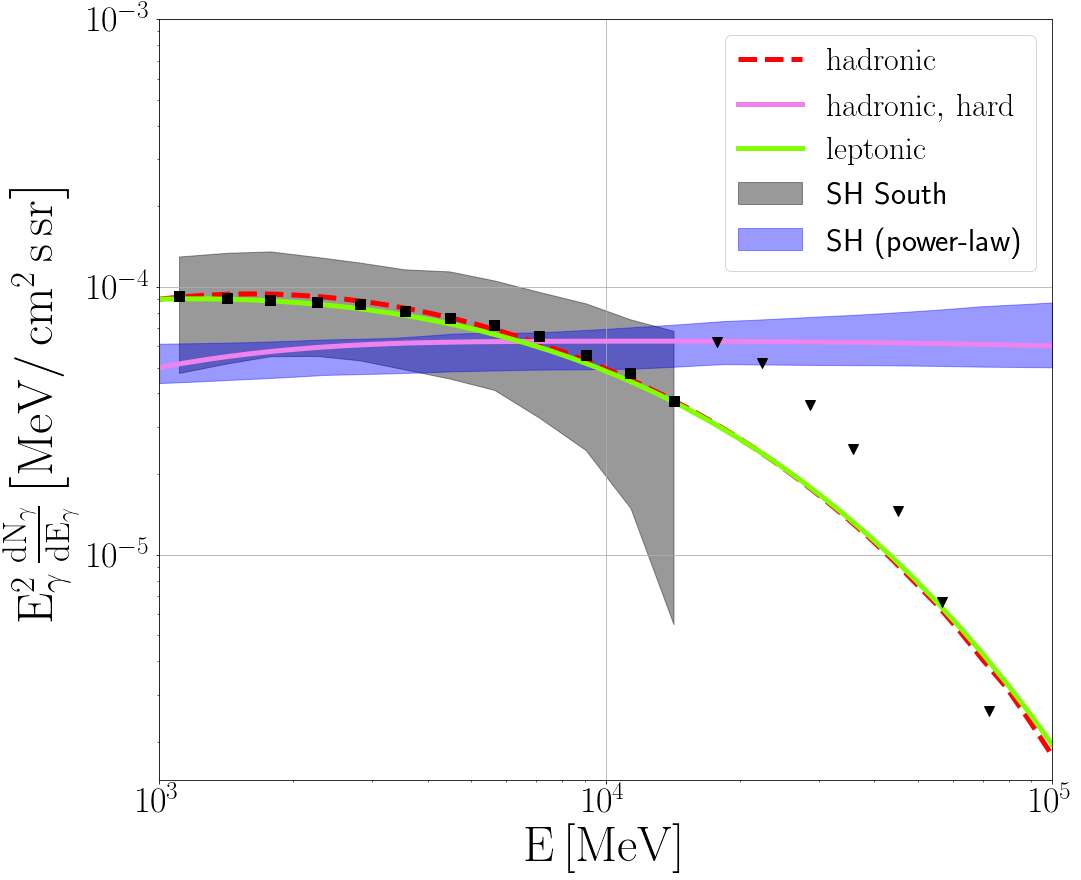}
    \caption{The black squares with the shaded region represent the the best fit to the $\gamma-$ray spectrum in the SH south, obtained by the analysis of \textit{Fermi}-LAT in~\protect\cite{Moskalenko-gamma-M31}. The shaded blue region is a power law fit to data (FSSC IEM analysis).
    The pink solid line represents the $\gamma-$ray flux in the case of the hard CR proton spectrum scenario considered in Sec.~\ref{sec:PeV}.
    The downward triangles are upper limits. The green solid (red dashed) line is a leptonic (hadronic) fit to data. 
    }
	\label{fig:fit-gamma}
\end{figure}

A qualitative fit to the $\gamma-$ray flux is shown in Fig.~\ref{fig:fit-gamma}, which has been obtained assuming a CR proton spectrum $\propto  E_{\rm GeV}^{-2.0}\, e^{-E_{\rm GeV}/110}$ or $\propto \rm E_{\rm GeV}^{-2.0}$ 
for the power law plus cutoff and pure power law scenario, respectively. The emission from p-p interactions has been computed following~\cite{Kamae-2008, Kelner-2006-pp}. 

If the CR proton luminosity $L_p$ is stationary over a time $\tau_{res}$, then the total energy in form of CR protons in the SH would be $L_p \times \tau_{res}$, which would result in an average CR energy density of the order of:
\begin{equation}
\omega_{CR,p} = \frac{L_p \tau_{res}}{V_{SH}} \sim 0.017 ~ n_{H,-3}^{-1} \rm eV/cm^3,
\end{equation}
where $V_{SH}$ is the volume of the SH region.
This is much smaller (about a factor of 50) than the typical energy density of CRs in the disk of the MW.

\section{Origin of the radiating particles}
\label{sec:origin}

As seen in the previous Section, the luminosity in form of energetic particles needed to explain the $\gamma$-ray halo of M31 is comparable to the CR luminosity of the MW. This is true for both the hadronic and the leptonic scenario (see Equations \ref{eq:Le} and \ref{eq:Lpp-SH}).
As M31 and the MW are two galaxies of the same type and of comparable masses, it is reasonable to assume that their CR energy outputs should also be of the same order. 

According to the standard model for CR origin~\citep{ReviewGabici-2019}, such energetic particles are accelerated at sources (maybe supernova remnants) located in the disk of the MW. The injection of CR particles in the interstellar medium is balanced by their escape from the Galaxy, which is due to a combination of spatial diffusion (random scattering in the turbulent galactic magnetic field) and advection (due to the presence of a galactic wind).   
Therefore, one might envisage a scenario where CRs (protons and/or electrons) would be accelerated at supernova remnants in the disk of M31 or injected from   the GC, and then transported into the extended halo where they would interact with ambient gas and/or radiation fields to produce the observed $\gamma$-ray emission.

However, such a scenario would not work. As seen in the previous Section, the problem with electrons is their very short radiative time scale, that would prevent them to be transported to large distances from the disk. For protons, the problem is related to their advective/diffusive transport from the disk/GC to the halo. Predictions from standard models involving a turbulent ambient magnetic field and a galactic wind invariably predict a decreasing intensity of CR protons for larger distances from the disk/GC (see Appendix for more details).
Thus, in order to fit the $\gamma$-ray flux observed from the SH, a very large intensity of CR protons must be present in the disk of M31, that should therefore be observed as a bright $\gamma$-ray source.
As the gaseous disk is not bright in $\gamma$-rays, we conclude that a CR origin of the $\gamma$-ray emission from the halo of M31 requires to go beyond the standard scenario for the production and transport of CRs in normal galaxies.

In this section we explore two scenarios for the production of the energetic particles responsible for the $\gamma$-ray emission from the SH. We consider first a model where CR protons are produced in the galactic center of M31, and are then transported into the halo by means of buoyant bubbles. In such scenario, the problem of a substantial decrease of the CR density with the distance from the GC can be overcome.
 
Then we consider a scenario where particles (either protons or electrons) are produced {\it in situ}, as a result of the acceleration at a gigantic shock located in the SH. Such a shock might be either an accretion shock, or the termination shock generated by a galactic outflow.

\subsection{Activity of the galactic center}
\label{sec:buoyant}

The discovery of the Fermi Bubbles in the MW is a spectacular signature of past nuclear activity in our Galaxy~\citep{Su-2010}.
They are two symmetric $\gamma$-ray emitting bubbles extending up to a distance of $\gtrsim 10$ kpc above and below the Galactic disk~\citep{ackermann2014}.
Very recently, the X-ray counterparts of Fermi Bubbles have been observed~\citep{predehl2020}.

Due to the similarity with the MW we will assume here that episodes of nuclear activity happens also in M31 (we remind that the existence in M31 of structures similar to the Fermi Bubbles has been proposed based on \textit{Fermi}-LAT observations~\citep{Pshirkov-2016}).
The origin of Fermi Bubbles is still debated. They could be inflated as the result of either intense star formation in the Galactic center (e.g.~\cite{crocker2011}), or accretion/ejection processes at the central supermassive black hole (e.g.~\cite{Guo-2012}).
The two scenarios involve mechanisms operating over different timescales (from few to few tens of Myr), and injecting energy at  different rates (from $\lesssim 10^{41}$ to $\approx 10^{43}$ erg/s), with overall energetics in the range spanning from $W_{B} \approx 10^{55}$ up to few times $10^{57}$ erg~\citep{Guo-2012,miller2016,barkov2014,yang2012}. 

Here, we investigate a scenario where CR protons are produced in recurring episodes of nuclear activity in M31, similar to that responsible for the creation of the Fermi Bubbles in the MW, and are then transported into the extended halo.
If $\nu_B = 10^{-2} \nu_{B,-2} \rm Myr^{-1}$ is the frequency of the episodes of nuclear activity in M31, the effective rate at which CR protons may be injected into the halo is 
\begin{equation}
L_p = \eta E_B \nu_B \sim 3.2 \times 10^{41} \eta E_{B,57} \nu_{B,-2} \rm erg/s ~,
\end{equation}
where $E_{B,57} = E_B/10^{57}$ erg and $\eta$ is an efficiency that takes into account the fact that only a fraction of the total energy involved in the process is converted into cosmic rays. The efficiency also accounts for possible adiabatic energy losses that particles may experience during the transport to the halo. After comparing this expression with the energy requirement in Eq.~\ref{eq:Lpp-SH} one gets:
\begin{equation}
\eta \approx 0.56 ~ \tau_{res,9}^{-1} n_{H,-3}^{-1} E_{B,57}^{-1} \nu_{B,-2}^{-1} 
\end{equation}
which is tight but not at all unfeasible.
For example, a moderate efficiency at the percent level could be achieved by considering a confinement time of CRs close to the age of the system ($\tau_{res,9} \sim 10$) and a typical energetic for a single episode of nuclear activity characterized by values of $E_{B,57}$ of the order of a few (comparable to the estimate made in~\cite{yang2012} for the energetic of Fermi Bubbles). 

The transport of CR protons from the galactic center to the halo must proceed in such a way to prevent particles to return to the disk and avoid in this way an overproduction of $\gamma$-rays there, due to proton-proton interactions in the interstellar gas.
This requirement might be accommodated, for example, by assuming that CR-inflated-buoyant-bubbles carry energetic particles to large distances from the disk, before being disrupted by plasma instabilities~\citep{Gull1973}. 
Such buoyant bubbles, often present in the central regions of clusters of galaxies~\citep{Jones-2005,Churazov-Bruggen-2001}, have been observed also in galaxies~\citep{finoguenov2008}.

Bubbles are typically found to rise at a buoyant velocity of the order of a fraction of the sound speed (which is $\sim 100$ km/s for the typical temperature of the hot diffuse circumgalactic gas~\citep{Zhang-2018}), while  their typical lifetime  could be as low as $\approx 10^8$ yrs~\citep{Churazov-Bruggen-2001, Jones-2005, Zhang-2018}, or significantly longer, even beyond $\sim 10^9$ yrs if the stabilizing action of a magnetic field is invoked (see~\cite{Zhang-2018} and references therein).
For a bubble lifetime of $\sim 10^9$ yrs and a sound speed of hundred km/s, every bubble would bring CRs up to a distance of $\lesssim 100$ kpc form the disk before releasing them in the halo~\citep{Jones-2005, Zhang-2018}. 

Once released in the galactic halo, CRs will spread diffusively to fill a region of $\approx 100$ kpc radius, as indicated by the extension of the $\gamma-$ray  emission in the SH. 
The time needed to fill such region depends on the CR diffusion coefficient $D$, which under most circumstances is an increasing function of particle energy. 

For particle energies for which the diffusion time $\tau_{res} \sim R_{SH}^2/(6~D)$ over a region of size $R_{SH} \sim 100$ kpc is shorter than the typical time between episodes of nuclear activity $1/\nu_B$, the contributions of several bubbles in the SH overlap and the CR population in that region and at those particle energies is stationary.
This happens for protons of energy smaller than $E_c$, when the diffusion coefficient is smaller than the critical value $D(E) < D(E_c) = 5\times 10^{30}\, R_{100}^2\nu_{B,-2}$ cm$^2$/s.
On the other hand, protons of energy larger than $E_c$ will populate the halo intermittently.
For this reason, a cutoff in the proton spectrum will occasionally appear, and could explain the power law plus cut-off fit to $\gamma-$ray data (black points in Fig.~\ref{fig:fit-gamma}), provided that $E_c \approx 100$ GeV.
Notice that the corresponding diffusion coefficient for $E = E_c$ would be somewhat larger than the value usually quoted for Galactic CR propagation at this energy~\citep{Strong-2007}. However, at the large distances from the disk considered here, an increase of the diffusion coefficient compared to the near-disk region may be indeed expected~\citep{Ptuskin-1997-wind, Strong-2007, Recchia-winds-I}.

A radically different scenario can be envisaged, if the diffusion coefficient is rather flat in energy, and/or if all CR protons injected during the entire lifetime of M31 are confined in the SH.
In this case, no cutoff is expected in the spectrum, that could extend well beyond the GeV domain.
This would be consistent with the pure power law fit to data  (shaded purple region in Fig.~\ref{fig:fit-gamma}).
We will explore the implications of this scenario in Sections~\ref{sec:PeV} and \ref{sec:multimessenger}.

We conclude this Section by noticing that the presence of a large scale galactic wind or outflow might prevent particles injected in the SH from coming back to the disk.
Such outflow could be the responsible for the presence of target material at $\sim 100$ kpc (see e.g.~\cite{Miller-2015, recchia2020}). 
Let us assume that CRs are released by a buoyant bubble at a distance $R_*$ from the disk.
It would take a time $t_* \approx R_*^2/6~D$ to spread over a region of size $R_*$.
For a galactic wind velocity $u_{out} = 10^3 u_{out,3}$ km/s~\citep{recchia2020} the condition of non-return to the disk is then $t_* \gtrsim R_*/u_{out}$.
This translates into another constrain on the particle diffusion coefficient: $D \lesssim 5 \times 10^{30} R_{*,2} u_{out,3}$ cm$^2$/s, where $R_{*} = 10^2 R_{*,2}$~kpc.

\subsection{{\it In situ} acceleration of electrons and protons}
\label{sec:acceleration}

As seen in Section~\ref{sec:leptonic}, a leptonic origin of the $\gamma$-ray emission from the SH requires an acceleration of electrons {\it in situ}.
Therefore, here we explore a scenario where CR electrons are accelerated at a giant strong shock located in the SH of M31. 

The maximum energy $E_{max,e}$ of accelerated electrons at a spherical shock of radius $R_s$ can be estimated by equating the acceleration time scale $\tau_{acc} \sim a ~D(E_e)/u_s^2$~\citep{drury1983} to the energy loss time $\tau_{loss} = E_e/({\rm d}E_e/{\rm d}t)$.
Here, $u_s = 10^3 u_{s,3}$ km/s is the velocity at which matter flows into the shock, ${\rm d}E_e/{\rm d}t$ is the inverse Compton scattering energy loss rate (see Eq.~\ref{eq:ICSlossrate}), and $a \approx 10$ is a numerical factor that depends on how much the shock compresses the gas and the magnetic field (see e.g.~\cite{Gaggero-2018}).
Assuming Bohm diffusion, $D = R_L(E_e) c/3$ with $R_L$ the particle Larmor radius, the resulting maximum energy is:
\begin{equation}
\label{eq:Emaxe1}
E_{max,e} \approx 34~ a_1^{-1} u_{s,3} B^{1/2}_{-6} \rm ~TeV
\end{equation}
where $a = 10~ a_1$.

The strength of the magnetic field in the halo of galaxies is a poorly constrained quantity. 
However, it is believed that particle acceleration at shocks is accompanied by an amplification of the field~\citep{Bell-2004}.
From observations of galactic supernova remnants, it has been inferred that a fraction $\xi_B \approx 3.5\%$ of the shock ram pressure is converted into magnetic field energy~\citep{Volk-2005}, i.e. $B_d^2/8\pi \, =\, \xi_B\, \rho_0\, u_s^2 $, where the subscript $d$ indicates that the magnetic field has been measured downstream of the shock, while $\varrho_0$ is the gas mass density of the intergalactic medium upstream of the shock.
By substituting this expression in Eq.~\ref{eq:Emaxe1} one finally gets:
\begin{equation}
E_{max,e} \approx 24 ~u_{s,3}^3 n_{0,-4}^{1/4} \left(\frac{\xi_B}{0.035} \right)^{1/4} \rm TeV    
\end{equation}
where $n_0 = 10^{-4} n_{0,-4}$ cm$^{-3}$ is the number density of the intergalactic gas.

Since the observed  $\gamma$-ray spectrum extends up to at least a photon energy $E_{\gamma} \approx 10$ GeV, the electron spectrum should extend at least up to an energy $E_{max,e} \approx 2$ TeV.
This conditions is satisfied if:
\begin{equation}
u_{s,3}  \gtrsim 0.43 ~ n_{0,-4}^{-1/12} \left(\frac{\xi_B}{0.035} \right)^{-1/12}
\end{equation}

This velocity is remarkably close to the free fall velocity at the edge of the SH:
\begin{equation}
v_{ff} \sim 0.29 \times 10^3 M_{12}^{1/2} R_{SH,2}^{-1/2} ~ \rm km/s    
\end{equation}
where $M = 10^{12} M_{12} M_{\odot}$ is the total mass of M31. 
We note, also, that the radius of the SH, $R_{SH} = 100 R_{SH,2}\, \rm kpc$, is of the same order of the virial radius of the system.
We may speculate, then, that particles are accelerated at a spherical accretion shock, that would process free falling intergalactic matter.
Energy would flow across the shock at a rate:
\begin{equation}
\label{eq:accretion}
L_s \approx (4 \pi R_{SH}^2) \frac{\varrho_0 v_{ff}^3}{2} \sim 3.4 \times 10^{42} R_{SH}^{1/2} n_{0,-4} M_{12}^{3/2} \rm erg/s
\end{equation}
which could very easily satisfy the energy requirement expressed by Eq.~\ref{eq:Le}.
While we notice that particle acceleration at accretion shocks is very often invoked around clusters of galaxies~\citep{Blasi-2007-clusters}, we should keep in mind that the very existence of accretion shocks around galaxies is debated~\citep{birnboim2003,ji2020}.

Another possibility is that the shock is formed as the result of the nuclear activity in the galactic center.
Following~\cite{cheng2011,miller2016} we may obtain a very rough estimate of the radius $R_s$ and expansion velocity $u_s$ of the shock by means of an analogy with wind blown bubbles. 
If $L_{GC} = 10^{43} L_{GC,43}$ erg/s is the time averaged rate of energy injection in the bubble due to galactic nuclear activity, we have~\citep{weaver1977}:
\begin{equation}
R_s \approx± \left( \frac{L_{GC}}{\varrho_0}\right)^{1/5} t^{3/5} ~~, ~~
u_s = \frac{3}{5} \frac{R_s}{t} ~~ .
\end{equation}
After setting $R_s \sim R_{SH}$ and $t \sim \tau_{GC}$, where $\tau_{GC} = 10^{9} \tau_{GC,9}$ yr overall duration of the nuclear activity (which may also consist of a number of recurrent episodes), we get:
\begin{equation}
u_s \approx 0.2 \times 10^3 ~ L_{GC,43}^{1/5} n_{0,-4}^{-1/5} \tau_{GC,9}^{-2/5} \rm km/s   
\end{equation}
where we normalized the energy injection rate to a very small fraction of the Eddington luminosity of the central supermassive black hole of mass $M_{BH}$~\citep{Ghisellini-2013}:
\begin{equation}
\label{eq:Ledd}
L_{Edd} \sim 1.3\times 10^{46} \left(\frac{M_{BH}}{10^8~M_{\odot}}\right) \rm erg/s ~ .
\end{equation}

In order to provide the CR electron luminosity needed to explain observations (Eq.~\ref{eq:Le}) the acceleration efficiency of electrons at the accretion shock must be of the order of $\eta_e \sim 5 \times 10^{-4} R_{SH}^{-1/2} n_{0,-4}^{-1} M_{12}^{-3/2}$.
With efficiency we mean here the fraction of the energy flowing across the shock which is converted into the CR electrons responsible for the $\gamma$-ray emission.
In the scenario where the shock is generated as a consequence of the galactic nuclear activity, a fraction $\eta_e \sim 2 \times 10^{-4} L_{GC,43}^{-1}$ of the power injected into the system has to be converted into CR electrons.

Besides electrons, also CR protons will be accelerated at the shock, most likely with a much larger efficiency, as inferred from the study of Galactic CRs~\citep{strongMW}.
The maximum energy of accelerated protons $E_{max,p}$ can be obtained, in this case, by equating the acceleration time $\tau_{acc}$ to the age of the shock $\tau_{s} = 10^9 \tau_{s,9}$ yr (which is of the order of $\tau_{age}$ for the accretion shock or $\tau_{GC}$ for the shock formed due to the nuclear activity), as proton-proton interactions are an extremely inefficient energy loss mechanism (see Eq.~\ref{eq:Tloss-pp}).

In fact, another condition should be satisfied in order to have acceleration up to the highest energies: particles must remain confined in the accelerator. This is equivalent to impose that the diffusion length of protons of energy $E_{max,p}$ should be smaller than the size of the accelerator ($\sim R_{SH}$). 
The two conditions result in very similar values for $E_{max,p}$, so we consider here the former, which gives:
\begin{equation}
E_{max,p} \approx 4.6 \times 10^2 ~ u_{s,3}^3 \tau_{s,9} n_{0,-4}^{1/2} \left(\frac{\xi_B}{0.035} \right)^{1/2} \rm PeV
\end{equation}
showing that the acceleration of CR protons can proceed up to very high energies, well beyond the $\sim$ 10-100 TeV needed to explain the $\gamma$-ray observations.
The $\gamma$-ray emission from the halo can be explained if the acceleration efficiency of protons at the accretion shock is of the order of $\eta_p \sim 0.01~ R_{SH}^{-1/2} n_{0,-4}^{-2} M_{12}^{3/2}$, where we assumed that CR protons remain trapped downstream of the shock for the entire lifetime of the system ($\tau_{res,9} \sim 10$).
In the galactic nuclear activity scenario, the efficiency would be $\eta_p \sim 5 \times 10^{-3} n_{0,-4}^{-1} L_{GC,43}^{-1}$.

We conclude this Section by noticing that so far we made the implicit assumption that the shock in the SH is strong and non radiative, i.e., characterised by a shock compression factor $r = 4$.
Given the low density and large temperature $T > 10^6$ K in the halo, the shock would become radiative in a time $\tau_{cool} \approx \epsilon/\Lambda \approx 1 ~ n_{0,-4}^{-1} (T/3 \times 10^6 {\rm K})^{3/2}$ Gyr , where $\epsilon$ is the thermal energy density of the gas and the Kahn approximation has been adopted for the cooling coefficient $\Lambda \approx 10^{-19} T^{-1/2}$ erg cm$^3$ s$^{-1}$~\citep{Cox-2005}.
Note that here $T$ is assumed to be the temperature downstream of the shock.
This implies that an hypothetical shock in the SH might become radiative. Even though it is often assumed in the literature that radiative shocks are not efficient particle accelerators, recent studies seems to suggest, instead, that acceleration efficiency at the percent level, similar to what is required in our scenario, are indeed possible~\citep{Steinberg-2018}.

\section{A giant CR halo around the MW and the origin of Icecube neutrinos}
\label{sec:PeV}

We now turn our attention to the MW.
The existence of a giant ($\sim 100-200$ kpc) CR halo surrounding our Galaxy was proposed in~\cite{Taylor-2014} as an explanation of the diffuse flux of multi-TeV neutrinos detected by Icecube~\citep{icecube2013}.  

The differential isotropic flux of astrophysical neutrinos (all flavors, neutrinos plus antineutrinos) measured at Earth can be fitted with a power law~\citep{Icecube-2020}:
\begin{equation}
\Phi_{\nu}^{IC}(E_{\nu}) \sim 6.37 \times 10^{-18} \left( \frac{E_{\nu}}{100~{\rm TeV}} \right)^{-2.87} \rm GeV^{-1} cm^{-2} s^{-1} sr^{-1}  
\end{equation}
The data reported in~\cite{Icecube-2020} refer to particle energies above $\approx 100$ TeV, and therefore the integrated isotropic flux is $F_{\nu}(>100~{\rm TeV}) \sim 1.2 \times 10^{-10}$ erg/cm$^2$/s/sr.
Assuming that the observed neutrinos are produced in the halo of the MW, at a typical distance of $R_H = 10^2 R_{H,2}$ kpc, then the differential neutrino emissivity from the entire MW is:
\begin{equation}\label{eq:MWQnu}
Q_{\nu}^{MW}(E_{\nu}) = (4 \pi)^2 \Phi_{\nu}^{IC}(E_{\nu}) R_H^2
\end{equation}
and the related luminosity can be estimated as:
\begin{equation}
L_{\nu}^{MW}(>100~{\rm TeV}) \approx 1.8 \times 10^{39} R_{H,2}^2 \rm erg/s ~ .
\end{equation}
If the neutrinos are the result of the decay of charged pions generated in proton-proton interactions in the circumgalactic gas, then an isotropic diffuse $\gamma$-ray emission above $\sim 100$ TeV  is also expected, with a luminosity comparable to that of neutrinos.

\begin{figure}[t]
	\centering
	\includegraphics[width=\columnwidth]{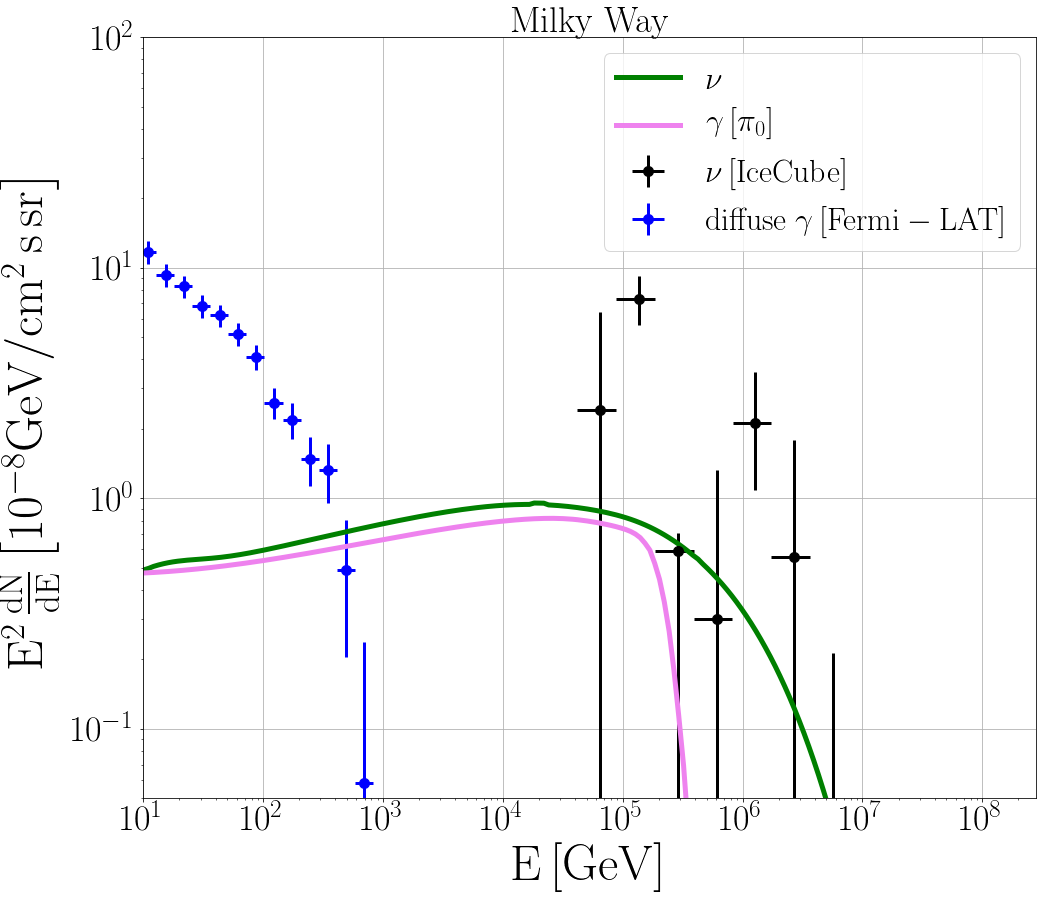}
    \caption{Isotropic diffuse neutrino and $\gamma$-ray emission observed by Icecube~\protect\citep{Icecube-2020} (black) and \textit{Fermi}-LAT~\protect\citep{Fermi-2015-diffuse} (blue data points). Solid lines are predictions for the neutrino (green) and $\gamma$-ray (pink) resulting from the interactions of CR protons with ambient gas in a $\sim 100$ kpc halo surrounding the MW.}
	\label{fig:MW}
\end{figure}

\begin{figure}[t]
	\centering
	\includegraphics[width=\columnwidth]{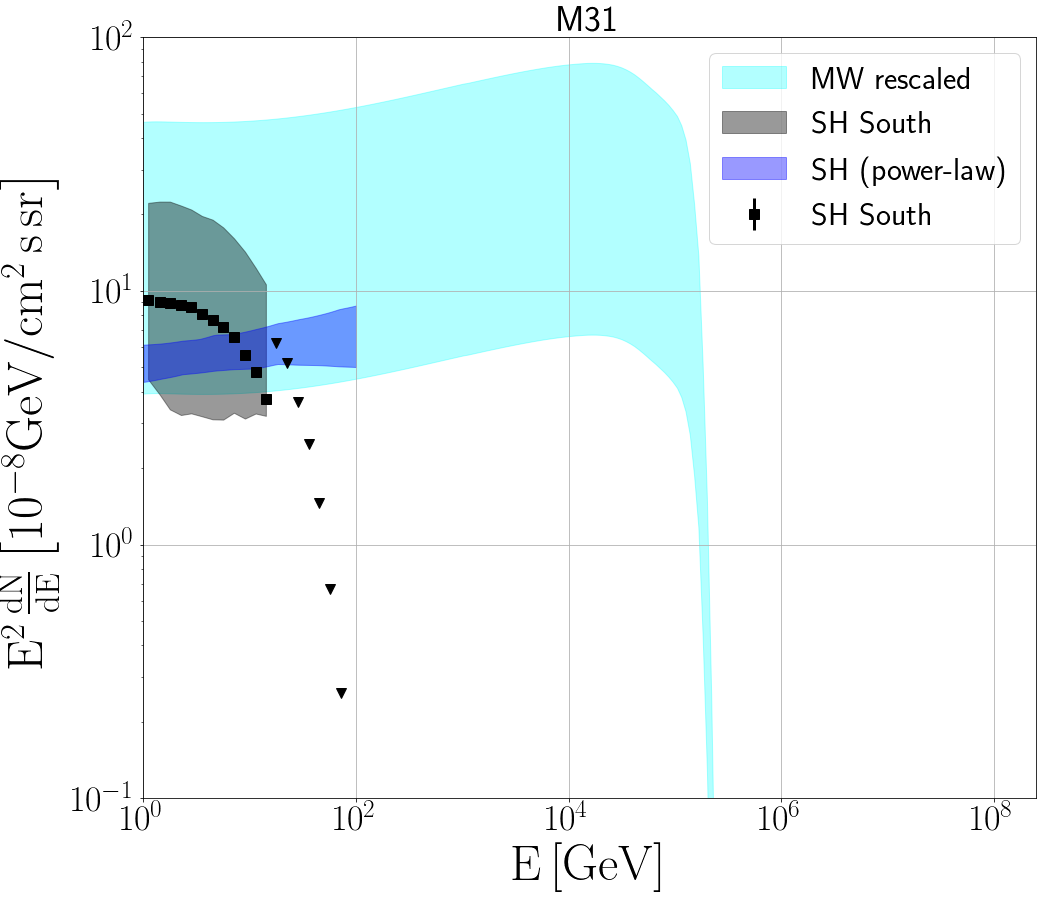}
    \caption{Expected $\gamma$-ray emission from M31 (shaded cyan region) obtained rescaling the predictions shown in Fig.~\protect\ref{fig:MW} for the halo of the Milky Way. The black data points and the shaded gray region show \textit{Fermi}-LAT observations of M31~\protect\citep{Moskalenko-gamma-M31}. 
    }
	\label{fig:MW-M31}
\end{figure}

The isotropic diffuse fluxes of $\gamma$-rays and neutrinos measured by \textit{Fermi}-LAT~\citep{Fermi-2015-diffuse} and Icecube~\citep{Icecube-2020}, respectively, are shown in Fig.~\ref{fig:MW}, together with the predictions from the $\gamma$-ray and neutrino emission from CR proton-proton interactions in the MW halo.
Predictions have been computed assuming a proton spectrum $\propto E_p^{-2} \exp(-E_p/20~{\rm PeV})$ normalized in such a way to contain a total energy equal to $W_p \sim 4.6 \times 10^{57} R_{H,2}^2 n_{H,-3}^{-1}$~erg.

The abrupt cutoff in the $\gamma$-ray spectrum reflects the fact that very high energy photons are absorbed due to pair production in the CMB~\citep{Lipari-2016}.
Absorption has been computed assuming photons traveled $100$ kpc.

It is interesting to note that such scenario would imply a recovery of the isotropic diffuse $\gamma$-ray emission above a photon energy of $E_{\gamma} \approx$ 1 TeV, at a level of $\sim 10^{-8}$ GeV/cm$^2$/s/sr, and extending up to $E_{\gamma} \lesssim$ 1 PeV. 
Available upper limits from observations performed at photon energies above $\sim 100$ TeV sit at about the same level~\citep{apel2017}.

It seems natural, at this point, to compare the two hadronic scenarios proposed above to explain the $\gamma$-ray and neutrino production in the halo of the MW and M31, respectively.
The observational constraints for the gas density in the halos of the two galaxies give very similar values (see e.g.~\cite{gupta2012,Bridge-MW-M31-2020}).
On the other hand, the CR proton content in the halo would scale linearly with the central SMBH mass in a scenario where the acceleration of particles is connected to galactic nuclear activity (see Eq.~\ref{eq:Ledd}) or with the total mass of the galaxy to the power 3/2 in an accretion scenario (see Eq.~\ref{eq:accretion}).
The ratio between the mass of M31 and of the MW is $\sim 2$~\citep{penarrubia2014}, while the ratio between the SMBH masses is $\sim 33$~\citep{Bender-2005}.
Therefore, assuming that the two galaxies can confine CRs for comparable times, one would expect the $\gamma$-ray and neutrino luminosities of M31 to exceed those of the MW by a factor of $\approx 2.8-33$.
Such rescaled fluxes are shown in Fig.~\ref{fig:MW-M31}, together with the spectrum of M31 observed by \textit{Fermi}-LAT for a power law plus exponential cutoff modeling of data (black points and shaded gray region) and for a power law model (blue line and shaded region)~\citep{Moskalenko-gamma-M31}.
The power law model derives from a less sophisticated analysis of \textit{Fermi}-LAT, but is claimed to be consistent with the more sophisticated approach. 
The good match between our prediction and \textit{Fermi}-LAT suggests a possible common origin of the GeV $\gamma$-ray emission from the halo of M31 and the neutrino emission from the halo of the Milky Way.
A firm assessment of the presence or not of a cutoff/steepening in the GeV $\gamma$-ray spectrum of M31 will be a crucial test for this scenario.

The large field of view and the superior sensitivity of the LHAASO 
detectors~\citep{LHAASO-2019} should be able to detect the multi-TeV $\gamma$-ray emission from M31, which implies that the scenario we propose will be tested in the near future.

Finally, another consequence of the similarity between M31 and MW is that, as pointed out already in~\cite{Taylor-2014}, multi-TeV CRs have to be confined in the galactic halos for long times, of the order of gigayears, to maintain the global energy budget at the levels discussed in Sec.~\ref{sec:energy}.

\section{Multiwavelength and multimessenger implications}
\label{sec:multimessenger}

In the scenario described above, the $\gamma$-ray emission observed from the halo of M31 is the result of CR proton-proton interactions with the circumgalactic gas.
If, as suggested in the previous Section, the spectrum of protons extends up to PeV energies, then the halo will also emit multi-TeV/PeV $\gamma$-rays and neutrinos.
However, it is straightforward to show that, according to the scenario illustrated in  Fig.~\ref{fig:MW-M31}, M31 should provide only a minor contribution (about $\approx 5$\%) to the total isotropic flux observed by Icecube (the number of neutrinos in the Icecube high-energy starting event sample considered here is 60~\citep{Icecube-2020}).
Moreover, neutrinos from M31 would come from a very extended region of apparent size (diameter) $\approx 15^{\circ} R_{H,2}$.

Assuming that all galaxies similar to the MW are surrounded by a giant CR halo, we can estimate their contribution to the isotropic diffuse neutrino flux.
If such a flux is in fact dominated by the emission coming from the MW halo, the contribution from other galaxies must be subdominant. 
This leads to the condition:
\begin{equation}
\label{eq:diffgal}
E_{\nu}^2 \Phi_{\nu}^{IC}(E_{\nu}) \gtrsim \frac{c \tau_{max}}{4 \pi} E_{\nu}^2 Q_{\nu}^{MW}(E_{\nu}) n_{gal}
\end{equation}
where we ignored the redshift evolution of sources, and assumed that only sources located within a distance $c \tau_{max}$ contribute to the diffuse flux. Here, $n_{gal} = 10^{-2} n_{gal,-2}$ Mpc$^{-3}$ is the density of galaxies in the local Universe.
The maximum distance $c \tau_{max}$ should be interpreted as follows: as it takes a long time, comparable to the age of the universe, to fill galactic halos with CRs, only old and massive galaxies in the nearby Universe are expected to be bright neutrino emitters. 
The constrain expressed by Eq.~\ref{eq:diffgal} is satisfied when:
$
(c \tau_{max}) \lesssim 3 ~ \rm Gpc
$ 
where we made use of Eq.~\ref{eq:MWQnu} and we set $n_{gal,-2} \sim 0.3$, which is an appropriate value for galaxies of mass comparable to that of the MW~\citep{blanton2009}.
We note that a very similar argument proposing that our Galaxy could be a typical emitter of high energy neutrinos and therefore could potentially provide a sizable fraction of the diffuse flux measured by Icecube was pushed forward in~\cite{gallo2018}.

The discovery potential of a point source for Icecube currently corresponds to a flux level of $\gtrsim 10^{-12}$ TeV/cm$^2$/s, if the source spectrum is $\propto E_{\nu}^{-2}$~\citep{aartsen2020}.
The halo of a galaxy like M31 would appear as point like (smaller than a degree) if located at distances larger than $\gtrsim$ 10 Mpc.
At that distance, the neutrino flux of a M31-like galaxy would be of the order of $\approx 2 \times 10^{-14}$ TeV/cm$^2$/s, a couple of orders of magnitude below the detection limit.
This implies that only galaxies characterised by an enhanced nuclear or starburst activity could be seen as neutrino sources.
This is particularly relevant in connection with the recent claim by the Icecube collaboration of a $\lesssim 3 \sigma$ excess of very high energy neutrinos from the galaxy NGC 1068, located at a distance of 14.4 Mpc~\citep{aartsen2020}. 
This galaxy exhibit both AGN and starburst activity, and therefore it seems to be an ideal neutrino emitter.
However, the non-detection of very high energy $\gamma$-rays from this object~\citep{aartsen2020} challenges an interpretation of the neutrino excess based on proton-proton interactions from the extended disk/halo region.
A possible combined explanation of $\gamma$-ray and neutrino observations of NGC 1068 was recently proposed in~\cite{Inue-2020}. 
Deeper observations in both neutrinos and $\gamma$-rays will help in clarifying this issue.

Besides the neutrino emission, also synchrotron emission in the X-ray domain is expected from galactic CR halos~\citep{Taylor-2014}.
Such emission would result from the secondary electrons produced in proton-proton interactions between CRs and the circumgalactic gas. 
As proton-proton interactions produce electrons and neutrinos of roughly the same energy, the spectrum of secondary electrons would extend up to $\approx$ PeV energies, which in turn implies that the synchrotron photons would reach an energy of the order of:
\begin{equation}
E_X \approx 20 ~ B_{H,\mu{\rm G}} E_{e,{\rm PeV}}^2 ~ \rm keV
\end{equation}
where $B_{H,\mu{\rm G}}$ is the magnetic field strength in the halo in $\mu$G, and $E_{e,{\rm PeV}} = E_e/{\rm PeV}$ the CR electron energy.

The maximum possible  X-ray flux is obtained when secondary electrons radiate synchrotron photons in the fast cooling regime, i.e., when the synchrotron energy loss time:
\begin{equation}
\tau_{syn}(E_e) \approx 10^4 E_{e,{\rm PeV}}^{-1} B_{H,\mu{\rm G}}^{-2} \rm yr  
\end{equation}
is shorter than all the other relevant time scales.
When this is the case, the peak synchrotron brightness is independent on the actual value of the magnetic field strength, and the halo of the MW would emit X-rays at a level 
$
\Phi^{MW}_X \approx 10^{-3} ~ \rm keV/cm^2/s/sr   
$,
which is much weaker than the extragalactic diffuse X-ray background~\citep{Gilli-2007}.
This would correspond to a peak global X-ray luminosity of the MW halo of the order of $L_X^{MW} \approx 2 \times 10^{37} R_{H,2}^2$ erg/s,
which implies that the halo of a galaxy similar to the MW located at a distance $d$ would be characterized by a peak X-ray flux and an angular extension equal to
$\approx 2 \times 10^{-14} R_{H,2}^2 (d/3~{\rm Mpc})^{-2}$ erg/cm$^2$/s and $\vartheta_H \sim 2^{\circ} R_{H,2} (d/3~{\rm Mpc})^{-1}$, respectively.

Even though the X-ray fluxes predicted above are quite large, one should keep in mind that the estimate done of the X-ray luminosity is most likely over-optimistic. 
This is because, in order to invoke the fast cooling regime, we implicitly assumed that the magnetic field strength in the galactic halo is of the order of at least few microgauss, which is the typical value found in the disk of galaxies.
A more appropriate value in the diluted halo might be a fraction of microgauss (comparable, for example, to the field strength measured in the intergalactic medium of rich clusters of galaxies, of density $\sim 10^{-3}$ cm$^{-3}$).
In this case, CR electrons would cool mainly through inverse Compton scattering off CMB photons, in a characteristic time $\tau_{ICS}$.
The estimate of the X-ray peak luminosity should be then corrected by a factor $\tau_{ICS}/\tau_{syn} < 1$, that for a magnetic field of 0.1 $\mu$G and for electron energies $E_e \approx 1$ PeV would be of the order of several times $10^{-2}$.
This would reduce X-ray fluxes to a level which is beyond the capabilities of current instruments.

\section{Conclusions}
\label{sec:concl}

The existence of a very extended ($\approx$ 100 kpc) gaseous halo around the MW has been revealed by a number of X-ray observations (see e.g.~\cite{gupta2012}).
If found to be a common feature of galaxies, gaseous halos might solve the problem of missing baryons in the Universe.
In~\cite{Taylor-2014}, it was proposed that, besides solving this problem, such gaseous halos could also shine in very high energy $\gamma$-rays and neutrinos, due to the interactions between CR protons and ambient gas. 
It was shown that, under certain assumptions on the CR luminosity of the MW, the neutrino emission from the halo could explain the diffuse flux of neutrinos measured by Icecube.

In this paper we investigated the observational consequences that the presence of a similar halo would have for M31, the closest massive galaxy to the MW.
We were motivated by the recent discovery in \textit{Fermi}-LAT of a giant $\gamma$-ray halo of size $\sim$ 100-200 kpc surrounding the galaxy.

Our main conclusion is that, provided CRs can be confined for long times (comparable to the age of the system) in the halos, both the isotropic diffuse neutrino emission observed by Icecube, and the extended $\gamma$-ray emission measured by \textit{Fermi}-LAT around M31 could be explained in terms of CR interactions with the circumgalactic gas.

We showed that such large halos may be explained through  CR protons produced in the galactic center of M31 and then transported into the halo by means of buoyant bubbles, or in a scenario where CRs, either electrons or protons, are accelerated \textit{in situ} at a large shock in the SH region. In the former case, the morphology of the emission from M31 is expected to be similar to that of Fermi Bubbles, but much more extended, while in the latter case, the emission is expected to be roughly spherically symmetric.

The time averaged luminosity of CRs to be injected in the halos is of the order of $\sim 10^{40}-10^{41}$ erg/s, comparable to the estimated luminosity of CR sources in the Galactic disk.
The scenario we propose is testable, as it predicts a multi-TeV $\gamma$-ray emission from the halo of M31 that is within the reach of instruments such as LHAASO.

If all galaxies are surrounded by gaseous halos, then they might all emit both $\gamma$-rays and neutrinos.
However, given the performances of current instruments, we estimated that the detection of galaxies located at distances significantly larger than M31 would be possible only in the presence of an enhanced nuclear or starburst activity, that could boost the acceleration of CRs.
A very recent claim from the Icecube collaboration on a $\lesssim 3 \sigma$ excess from the direction of NGC 1068, a Seyfert type galaxy exhibiting starburst activity and located at a distance of 14.4 Mpc, might fit with this prediction. On the other hand, the lack of $\gamma$-ray emission from that object poses problems to any interpretation of data based on CR interactions with the gas in an extended disk/halo system.

We conclude by recalling that an alternative and viable scenario to interpret the observations of M31 could involve leptonic interactions (namely, inverse Compton scattering) operating in its halo.
Of course, in this case no neutrinos would be expected from M31.

\acknowledgments

 The authors would like to thank M. Brueggen for helpful discussions.
 SR and SG acknowledge support from the region \^{I}le-de-France under the DIM-ACAV programme, from the Agence Nationale de la Recherche (grant ANR- 17-CE31-0014), and from the Observatory of Paris (Action F\'ed\'eratrice CTA). This project has received funding from the European Union's Horizon 2020 research and innovation programme under the Marie Sk\l{}odowska-Curie grant agreement No. 843418 (nuHEDGE).

%




\appendix

\section{Standard models of cosmic ray propagation}
\label{sec:standard-prop}

The extended $\gamma-$ray emission of M31 is very hard to be accounted for in the standard scenario in which CRs are generated by sources, e.g SNRs, located in the galactic disk and in typical scenarios of CR propagation from the disk. 
This is obvious for CR electrons, which would not be able to travel $\sim 100$ kpc even considering only losses  due to  ICS on the CMB. As for CR protons, we show that, in any common transport  scenario the CR density inevitably decreases significantly with the distance from the disk. This is at odds with the detected $\sim 100$ kpc extended emission.

In fact, in typical propagation models,   CRs are expected to diffuse away from the disk while being  scattered on plasma waves. Such waves could be injected in disk by astrophysical sources~\citep{Evoli-2018-halo} or could be produced by the CR streaming instability~\citep{Kulsrud-1969,Skilling-1975-II, Blasi-2019-review-self-conf}. However, in both cases, the magnetic turbulence is expected to decrease away from the disk, resulting in  a CR diffusion coefficient which tends to increase with the distance from the disk. In such scenario, CRs are expected to free stream above few kpc from the disk~\citep{Blasi-2013-review, Amato-2014-review}. Evidently all this would lead to a (fast) decrease of the CR density toward the outer halo.

Also in more complex models,  involving, together with diffusion, the possible presence of a galactic wind or breeze which advect CRs  (and which could be responsible for the presence of target material at $\sim 100-200$ kpc from the disk of Andromeda) this conclusion does not change. 
In fact, following the approach by~\cite{Breitschwerdt-1991-I, Everett-2008, Recchia-winds-I, Recchia-winds-II}, in a stationary wind/breeze model, characterized by the flux tube area $A(z)$, with $z$ the distance from the disk, the gas $n(z)$ density and velocity $ u(z)$, and the CR pressure $ P_c(z)$ are related by conservation laws
\begin{align}
  & n\,u\,A\,= \, \rm const\\ \nonumber
  & P_c\,(u\,A)^{\gamma_c} = \rm  const,
\end{align}
where $\gamma_c$ is the adiabatic index of the CR gas ($\gamma_c = 4/3$ for relativistic particles). 
At large enough $z$, the flux tube is expected to open up spherically and the gas density is  observed to decrease with $z$~\citep{Miller-2013, Miller-2015}. Moreover, at large $z$ the wind velocity becomes constant, while it decrease with $z$ in the case of breeze. We get
\begin{align}
 & \rm A(z)\, \propto \, z^2\\ \nonumber
 & \rm u(z)\, \propto \, z^{-\beta}\\ \nonumber
 & \rm n(z)\, \propto \, z^{\beta-2}\\ \nonumber
 & \rm P_c(z)\, \propto \, z^{(\beta-2)\gamma_c}.
\end{align}
Here $ \beta < 2$ (otherwise $n$ would increase with $z$) and $\beta = 0$ for a wind.
Thus, in both cases the CR pressure would decrease with $z$. 
In addition, \cite{Recchia-winds-I} showed that in a wind the CR spectrum tends to become progressively harder with $z$, thus enhancing the decrease of the CR density at energies below $\sim 1000$ GeV.\\

Recently, a interesting self-confinement scenario has been  put forward by~\cite{Blasi-2019, Blasi-2019-review-self-conf} in the case of the MW. 
The idea is that  the CR current of CRs produced in the disk and trying to free stream to infinity in  a progressively smaller background magnetic field,  leads to the excitation of a non resonant plasma instability, the Bell instability. 
As far as the CR propagation perpendicular to the disk can be considered as one-dimensional, the instability induces  a strong amplification of the background magnetic field and a suppression of the CR diffusion coefficient. This happens for distances  from the disk smaller than roughly the disk radius $ R_d\,$($\sim 10$ kpc for both M31 and the MW).
The instability is excited on scales much smaller than the Larmor radius of CRs and saturates when the scale  becomes comparable to the Larmor radius. 
In addition, the resulting large CR pressure gradient leads to a displacement of the background medium at a speed which is roughly at the level of the Alfv\'en speed computed in the amplified magnetic field. 
The resulting confinement time in a region of  $\simeq 10$ kpc can be as large as $\sim 10^{8}-10^{9}$ yrs.\\
The instability is excited only if the background magnetic field is smaller than the saturation value
\begin{equation}\label{eq:Bsat}
   B_{sat} \approx 2.2 \times 10^{-8}\; L_{41}^{1/2}\; r_{10}^{-1}\; \rm G,
\end{equation}
where $ L_{41}$ is the CR luminosity in units of $10^{41} $ erg/s and $ r_{10}$ is the size of the CR source region (e.g the disk) in units of 10 kpc.
CR diffusion is expected to proceed at the Bohm limit in the saturated magnetic field and CR will experience ad  advection velocity, respectively, given by:
\begin{align}
  &  D(E) \approx 1.5\times 10^{24}\, E_{GeV}\, L_{41}^{-1/2}\, r_{10}\, \rm cm^2 /s \\ \label{eq:D-self-conf} \nonumber
  & \Tilde{v}_A \approx 5\, L_{41}^{1/2}\, r_{10}^{-1}\, n_{-4}^{-1/2}\, \rm   km/s.
\end{align}
However, at distances larger than $\sim R_d$, the one-dimension propagation assumption is no more valid, the CR density tends to decrease spherically and the self-confinement becomes quickly inefficient~\citep{Blasi-2019, Blasi-2019-review-self-conf}. 
Thus, even in a scenario of strong CR confinement, the CR density is expected to rapidly drop with the distance from the disk.


\bibliography{biblio}{}
\bibliographystyle{aasjournal}



\end{document}